\documentclass[fleqn,12pt]{article}
\usepackage[numbers,sort&compress]{natbib}
\usepackage[footnotesep=0.4in]{geometry}
\usepackage{graphicx}
\usepackage{caption,subcaption}
\usepackage{amsmath}
\usepackage{bbding}
\usepackage{graphics}
\usepackage{pifont}
\usepackage{amsfonts}
\usepackage{float}
\makeatletter
\newcommand\figcaption{\def\@captype{figure}\caption}
\newcommand\tabcaption{\def\@captype{table}\caption}
\begin{document}
	
	\date{}
	\title{Periodic parabola solitons for the nonautonomous KP equation}
	\author{Yingyou Ma$^{1}$\,,\,Zhiqiang Chen$^{1}$\,,\, Xin Yu$^{ 2, }$\thanks{Corresponding
			author, with e-mail address as yuxin@buaa.edu.cn}
		\\
		\\{\em  $^{1}$School of Physics, Beihang University, Beijing 100191, China}\\
		{\em  $^{2}$Ministry-of-Education Key Laboratory of Fluid Mechanics and
			National}\\
		{\em  Laboratory for Computational Fluid Dynamics,  Beihang University, }\\
		{\em   Beijing 100191, China} \\
				 \\
	}\maketitle

\vspace{8mm}

\begin{abstract}
	Kadomtsev-Petviashvili (KP) equation, who can describe different models in fluids and plasmas, has drawn investigation for its solitonic solutions with various methods. In this paper, we focus on the periodic parabola solitons for the (2+1) dimensional nonautonomous KP equations where the necessary constraints of the parameters are figured out. With  Painlev\'{e} analysis and Hirota bilinear method, we find that the solution has six undetermined parameters as well as analyze the features of some typical cases of the solutions. Based on the constructed solutions, the conditions of their convergence are also discussed.	
\end{abstract}

\vspace{3mm}


\vspace{3mm}


\vspace{20mm}

\newpage
\noindent {\Large{\bf I. Introduction}}

\vspace{3mm}In several aspects of physics, some dynamical systems can be described by nonlinear partial differential equations (PDEs)~\cite{Ablowitz1991Solitons}. While investigating them, we pay attention to some soliton solutions for their significance both in theoretical and practical values . The Kadomtsev-Petviashvili (KP) equation, as follows:
\begin{equation}\label{cequation}
\hspace{10mm}(u_t\,+6\,u\,u_{x}+u_{xxx})_x+3\, \sigma \,u_{yy}=0\,,\quad \sigma =\pm 1\,,
\end{equation}
is such a nonlinear PDE which can describe surface wave with low amplitude~\cite{Kadomtsev1970On}. Several researches focusing on its solutions have emerged including algebraically decaying solutions~\cite{1978JMP....19.2180A}, lump solutions~\cite{Satsuma1979Two}, rogue waves~\cite{Waseda2013Rogue} and periodic solitons~\cite{Liu2010New}. For more complicated models, such as the ones considered the variation of depth and density, nonautonomous KP equation with variable coefficients should be investigated\cite{David1987Integrable,David1989Solitons} and some researches have been finished~\cite{Zhu1993Lax,TIAN1997125,2006CoTPh..46..793Z,L2010Explicit,Li2010Painlev, Yu2016Parabola,Liang2011Transformations,Tian2006Transformations,Yomba2004Construction}. In this paper, we set the nonautonomous KP equation in this form:
\begin{equation}\label{equation}
\hspace{10mm}[u_t\,+a(t)\,u\,u_{x}+b(t)\,u_{xxx}]_x+\sigma\,c(t)\,u_{yy}+d(t)\,u_{xy}+[e_1(y,t)+e_2(t)x]\,u_{xx}+f(t)u_{x}=0\,,
\end{equation}
where $x$ and $y$ are scaled space coordinates, $t$ is scaled time
coordinate, $a(t)$, $b(t)$, $c(t)$, $d(t)$, $e_1(y,t)$, $e_2(t)$ and
$f(t)$ are inhomogeneous coefficients while $b(t)$ and $c(t)$ are both positive and $\sigma=\pm1$. In the following, Eq.~(\ref{equation}) with $\sigma=+1$ and $\sigma=-1$ will be named KP-I and KP-II equation, respectively. The convergence and interactions of parabola exponent solitons have been already investigated for KP-I equation~\cite{Yu2016Parabola}, but it remains unknown when trigonometric function is considered. In this case, the features and convergence of the solutions will become more complicated where the difference of KP-I and KP-II equation can be distinct. Such characteristics are drawing more and more attention in fluid and plasma physics, especially the singularities, which may appeal to some novel physics phenomena~\cite{2006CoTPh..46..793Z}. 

\vspace{10mm}

\noindent {\Large{\bf II. Periodic parabola solitons}}

\vspace{3mm} The  Painlev\'{e} analysis~\cite{Weiss} for KP-I equation has been finished in Ref.~\cite{Yu2016Parabola}, whose $c(t)$ should be replaced by $\sigma c(t)$ somewhere when KP-II equation is taken into consideration. The integrable conditions are:
\begin{equation}\label{cc1}
\hspace{10mm}a(t)=6\,\rho\,b(t)^{\frac{3}{4}}\,c(t) ^{\frac{1}{4}}\,e^{\int{\![f(t)-2e_2(t)]dt}}\,,
\end{equation}
\begin{align}\label{cc2}
&\hspace{10mm}e_1(y,t)=-\frac{e_2(t)^2 y^2}{2 \sigma c(t)}-\frac{3 b'(t)^2
	y^2}{16 b(t)^2 \sigma c(t)}+\frac{3 c'(t)^2 y^2}{16
	\sigma c(t)^3}+\frac{3 e_2(t) b'(t) y^2}{8 b(t) \sigma
	c(t)}\notag\\
&\hspace{26mm}+\frac{e_2(t) \sigma c'(t) y^2}{8
	c(t)^2}-\frac{e_2'(t) y^2}{2 \sigma c(t)}+\frac{b''(t)
	y^2}{8 b(t) \sigma c(t)}-\frac{\sigma c''(t) y^2}{8
	c(t)^2}+\alpha _2(t) y+\alpha _1(t)\,,
\end{align}
where $\rho$ is a nonzero constant, $\alpha _1(t)$ and $\alpha
_2(t)$ are introduced arbitrary functions of $t$, and $'$ denotes
the derivative with respect to $t$.

In this paper, we propose the similar generalized dependent
variable transformation with Ref.~\cite{Yu2016Parabola},
\begin{equation}\label{bianhuan}
\hspace{10mm}u\,=\frac{2}{\rho
}\,b(t)^{\frac{1}{4}}\,c(t)^{-\frac{1}{4}}\,\,e^{-\int{\![f(t)-2e_2(t)]dt}}({\rm
	log} \Phi)_{xx}
+\Psi(x,y,t)\,,\\
\end{equation}
\begin{equation}
\hspace{10mm}\Psi(x,y,t)=\Psi_{1}(t)x+\Psi_{2}(y,t)\,,
\end{equation}
\begin{equation}
\hspace{10mm}\Psi_{1}(t)=\frac{1}{a(t)} \Big [ \beta_{3}
(t)-e_2(t)+\frac{b'(t)}{4 b(t)}-\frac{c'(t)}{4
	c(t)}\Big]\,,
\end{equation}
\begin{equation}\label{Phi2}
\hspace{10mm}\Psi_{2}(y,t)=-\frac{a(t) \Psi _1(t)^2 y^2}{2
	\sigma c(t)}-\frac{f(t)
	\Psi _1(t) y^2}{2 \sigma c(t)}-\frac{\Psi _1'(t) y^2}{2
	\sigma c(t)}+\beta _2(t) y+\beta _1(t)\,,
\end{equation}
where $\Phi$ is a function of $x$, $y$ and $t$, $\beta _1(t)$,
$\beta _2(t)$ and $\beta _3(t)$ are arbitrary functions of $t$. Under the conditions~(\ref{cc1})-(\ref{Phi2}), Eq.~(\ref{equation}) can be
transformed into its bilinear form as below,
\begin{equation}\label{bilinear}
\hspace{10mm}\Big[D_xD_t+b(t)\,D_{x}^{4}+\sigma c(t)\,D_{y}^{2}+d(t)\,D_{x}\,D_{y}+\varphi_{1}(x,y,t)
\,D_x^{2}+\varphi_2(t)\frac{\partial}{\partial x}\Big] \Phi \cdot
\Phi=0\,,
\end{equation}
where
\begin{align}
&\hspace{10mm}\notag \varphi_{1}(x,y,t)=\alpha _2(t) y+\alpha _1(t)+
6\,\rho\,b(t)^{\frac{3}{4}}
\,c(t)^{\frac{1}{4}}\,e^{\int{\![f(t)-2e_2(t)]dt}} \Big [\beta _2(t) y+\beta _1(t)\Big ]
\\&\hspace{10mm}+\frac{x b'(t)}{4
	b(t)}-\frac{x c'(t)}{4 c(t)}-\frac{\beta_{3} (t)^2 y^2}{2\sigma c(t)}+\frac{\beta_{3} (t) b'(t)
	y^2}{8 b(t) \sigma c(t)}+\frac{3 \beta_{3} (t)\sigma  c'(t) y^2}{8
	c(t)^2}-\frac{\beta_{3} '(t) y^2}{2 \sigma c(t)}+x \beta_{3} (t)\,,
\end{align}
\begin{equation}
\hspace{10mm}\varphi_2(t)=\frac{b'(t)}{4 b(t)}-\frac{c'(t)}{4
	c(t)}\,,
\end{equation}
\begin{equation}
\hspace{10mm}\frac{\partial}{\partial x} \Phi \cdot \Phi=2 \Phi
\Phi_x\,,
\end{equation}
and $D_{x}^{m}D_{t}^{n}$ is the Hirota bilinear derivative
operator~\cite{Hirota2004The,PhysRevLett.27.1192} defined by
\begin{eqnarray}
\hspace{0mm}D_{x}^{m}D_{y}^{n}D_{t}^{p} a\cdot
b\equiv\left(\frac{\partial}{\partial x}-\frac{\partial}{\partial
	x^{'}}\right)^{m}\,\left(\frac{\partial}{\partial
	y}-\frac{\partial}{\partial
	y^{'}}\right)^{n}\,\left(\frac{\partial}{\partial
	t}-\frac{\partial}{\partial
	t^{'}}\right)^{p}\,a(x,y,t)\,b(x^{'},y^{'},t^{'}) \bigg
|_{x^{'}=x,\,y^{'}=y,\,t^{'}=t}\,.\
\end{eqnarray}
Similar to the periodic linear soliton solutions in Ref.~\cite{PhysRevLett.27.1192}, the solution of Eq.~(\ref{bilinear}) can be set periodic
parabola solitonic as below (without loss of generality, we assume $b_2 >0$):
\begin{align}\label{Phi}
&\hspace{10mm}\notag \Phi=b_1\,e^{k_1(t)\,x+l_{11}(t)\,y+l_{12}(t)\,y^2+w_1(t)}+b_2\,\cos\left[k_2(t)\,x+l_{21}(t)\,y+l_{22}(t)\,y^2+w_2(t)\right]
\\&\hspace{13mm}+b_3\,e^{-[k_1(t)\,x+l_{11}(t)\,y+l_{12}(t)\,y^2+w_1(t)]}\,.
\end{align}
Substituting Eq.~\ref{Phi} into Eq.~\ref{bilinear}, we can derive an equation consisting of different terms, whose coefficient should be equaled to 0 due to the arbitrary coordinates. Such equations yield eight explicit constraints and one implicit constraint. The explicit ones are ($i=1,2$):
\begin{equation}\label{boshu}
\hspace{10mm}k_i(t)=C_{ki}\,b(t)^{-\frac{1}{4}}\,c(t)^{\frac{1}{4}}\,e^{-\int
	\beta_{3} (t)\, dt}\,,
\end{equation}
\begin{equation}\label{yfangxiang1}
\hspace{10mm}l_{i2}(t)=\frac{1}{2}\sigma\,C_{ki}\,b(t)^{-\frac{1}{4}}\,c(t)^{-\frac{3}{4}}\,\beta_{3}
(t)\,e^{-\int \beta_{3} (t)\, dt}\,.
\end{equation}
To simplify our process, we introduce three new functions $m(t)$, $n_1(t)$ and $n_2(t)$. They are defined as
\begin{align*}
&\hspace{10mm}m(t)\notag=\int b(t)^{-\frac{1}{4}}\,c(t)^{-\frac{3}{4}}\, e^{\int \beta_3 (t) \, dt} \Big[6 \, \rho \, b(t)^{\frac{3}{4}} \, c(t)^{\frac{5}{4}} \,\Lambda _2(t) \, e^{\int \left(f(t)-2 e_2(t)\right) \, dt}  \\
&\hspace{23mm}+c(t) \alpha _{10}(t)+\sigma d(t) \beta_3 (t)\Big]\,dt \,, \\
&\hspace{10mm}n_1(t)=b(t)^{\frac{1}{4}} c(t)^{\frac{3}{4}} \bigg[-C_{{k1}}
\left(-3 C_{ {k2}}^4+\sigma C_{ {l1}}^2-\sigma C_{ {l2}}^2\right)-2\sigma
C_{ {l1}} m(t)
\left(C_{ {k1}}^2+C_{ {k2}}^2\right)\notag \\
&\hspace{23mm}-\sigma C_{ {k1}} m(t)^2
\left(C_{ {k1}}^2+C_{ {k2}}^2\right)+2 C_{{k1}}^3
C_{{k2}}^2-C_{{k1}}^5-2\sigma C_{{k2}} C_{{l1}}
C_{{l2}}\bigg]\,,\\
&\hspace{10mm}n_2(t)=b(t)^{\frac{1}{4}} c(t)^{\frac{3}{4}} \bigg[-C_{{k2}}
\left(-3 C_{ {k1}}^4+\sigma C_{ {l1}}^2-\sigma C_{ {l2}}^2\right)-2\sigma
C_{ {l2}} m(t)
\left(C_{ {k1}}^2+C_{ {k2}}^2\right)\notag \\
&\hspace{23mm}-\sigma C_{ {k2}} m(t)^2
\left(C_{ {k1}}^2+C_{ {k2}}^2\right)+2 C_{{k2}}^3
C_{{k1}}^2-C_{{k2}}^5-2\sigma C_{{k1}} C_{{l1}}
C_{{l2}}\bigg]\,.
\end{align*}
Then we have
\begin{equation}\label{yfangxiang2}
\hspace{10mm}l_{1i}(t)=e^{-2 \int \beta_{3} (t) \, dt}\left( C_{li}-C_{ki}m(t)\right)\,,
\end{equation}
\begin{align}\label{sesan}
&\hspace{5mm}w_i(t)=\int e^{-3 \int \beta_3 (t) \, dt}\bigg[\frac{1}{C_{{k1}}^2+C_{{k2}}^2}n_i(t)-6 \rho  b(t)^{\frac{1}{2}} c(t)^{\frac{1}{2}}
C_{ki} \Lambda _1(t) e^{\int (f(t)-2
	e_2(t)+2\beta_3 (t)) \,
	dt} \notag \\
&\hspace{18mm}-b(t)^{-\frac{1}{4}}c(t)^{\frac{1}{4}} C_{{ki}} \alpha _9(t) e^{ \int 2\beta_3 (t) \,
	dt}+d(t) e^{\int \beta_3 (t) \, dt}
\left(C_{{ki}} m(t)-C_{{li}}\right)\bigg]\,dt\,.
\end{align}
 Besides, the implicit constraint is:
\begin{equation}\label{CON}
3 \left(C_{k1}^2+C_{k2}^2\right){}^2 \left(4 b_1 b_3 C_{k1}^2+b_2^2
C_{k2}^2\right)- \sigma \left(b_2^2-4 b_1 b_3\right) \left(C_{k2} C_{l1}-C_{k1}
C_{l2}\right){}^2=0\,,
\end{equation}
which is the key to determine the convergence.\\

In one word, the independence of $k_1(t),k_2(t),l_{11}(t),l_{12}(t),l_{21}(t),l_{22}(t),w_1(t),w_2(t),b_1,b_2,b—_3$ can be changed to be any six ones of seven parameters  $C_{k1},C_{k2},C_{l1},C_{l2},b_1,b_2, b_3$.

\vspace{3mm}
\noindent {\Large{\bf III. Solutions' convergence}}

\vspace{3mm}Here we set the moving characteristic line as $g_1$ ans $g_2$ for simplification:
\begin{align*}
&\hspace{10mm}g_1=k_1(t)\,x+l_{11}(t)\,y+l_{12}(t)\,y^2+w_1(t)\,,\\
&\hspace{10mm}g_2=k_2(t)\,x+l_{21}(t)\,y+l_{22}(t)\,y^2+w_2(t)\,,\\
&\hspace{10mm}\Phi=b_1 e^{g_1}+b_2\cos g_2 +b_3 e^{-g_1}\,.
\end{align*}
Since there is a term of $( \text{log}\Phi)_{xx} $ in the solutions, $u$ will be infinite in some area when $\Phi=0$. Thus, we want to find the maximum or the minimum value of $\Phi$ in different situations. In the following discussion, $h_1$ and $h_2$ represent the extreme line for $g_1$ and $g_2$, respectively.

\paragraph{(a)} $b_1>0, b_3>0$ and $k_1(t)k_2(t)\ne0$. In this case, the characteristic line will be parabolic and $\Phi$ must be positive in some area. The minimum of $b_2 \cos g_2$ will be $-b_2$ when $h_2=(2n+1)\pi$, $n\in N$.  Therefore, $b_2 \cos g_2$ will take the minimum value $-b_2$ in a set consisted of an infinite number of  uniformly spaced parabolas $\{h_2\}$. Similarly, $b_1 e^{g_1}+b_3 e^{-g_1}$ will take the minimum value $2\sqrt{b_1 b_3}$ in a parabola: $h_1=\frac{1}{2}\ln\frac{b_3}{b_1}$. We can modify the ratio between $b_1$ and $b_3$ to shift $h_1$. However, unless $h_1$ and $h_2$ is just the same characteristic line, it's obvious that the parabola $h_1=\frac{1}{2}\ln\frac{b_3}{b_1}$ will intersect with the set $\{h_2\}$ of $h_2=(2n+1)\pi$ in some area, where $\Phi$ take the minimum value $2\sqrt{b_1 b_3}-b_2$. \\

With the restrict of Eq.~(\ref{CON}), it's easy to prove $\sigma\left( b_2^2-4b_1b_3\right) >0$. Thus, if $\sigma=1$, $2\sqrt{b_1 b_3}-b_2<0$. In one word, for continuous function $\Phi$, there must be some place where $\Phi=0$, leading to the conclusion that $u$ will not always limited in this condition for KP-I unless $g_1$ and $g_2$ is the same characteristic line. To the opposite, when $\sigma=-1$, which means $2\sqrt{b_1 b_3}-b_2>0$, $\Phi$ is positive everywhere so $u$ obtains convergence for KP-II.\\

For the case when $g_1$ and $g_2$ is the same characteristic line (which means $C_{k1}/C_{k2}=C_{l1}/C_{l2}$), due to Eq.~(\ref{CON}), we can derive $b_1 b_3<0$, which leads to (d).

\paragraph{(b)} $b_1<0, b_3<0$ and $k_1(t)k_2(t)\ne0$. This is similar to Case 1. Here $b_2-2\sqrt{b_1 b_3}$ will be the maximum value $\Phi$ takes and  $\Phi$ must be negative some place. Like Case 1, for KP-I $b_2-2\sqrt{b_1 b_3}>0$, then $\Phi$ will have zero point making $u$ unlimited. On the contrary, for KP-II the maximum value is still below zero so $u$ has a bound.

\paragraph{(c)}$b_1 b_3 >0$ while $k_1(t)k_2(t)=0$. In this case, one of the characteristic line will turn to be a straight line but the conclusion from Case 1 and 2 will not change because a parabola will always intersect with a set of infinite uniformly-spaced straight lines and a straight line will always intersect with a set of infinite uniformly-spaced parabolas or straight lines.

\paragraph{(d)}$b_1 b_3 \le 0$. Here the high slope of exponent function will definitely bring $\Phi$ zero points, for both KP-I and KP-II.\\

In one word, if $b_1 b_3 \le 0$, both KP-I and KP-II will be unlimited in some places. If $b_1 b_3 >0$, only KP-II can keep the convergence when the shapes of the two characteristic lines are different.

\vspace{3mm}
\noindent {\Large{\bf IV. Different solutions}}

\vspace{3mm}
When we solve Eq.~(\ref{CON}) with symbolic computation,  the solutions can be classified by 5 cases depending on the characteristic line and whether $b_3$ equals to 0. In the discussion and figures, we set $\beta_3(t)=e_2(t)-\frac{b^{'}(t)}{4b(t)}+\frac{c^{'}(t)}{4c(t)}$ and $\beta_1(t)=\beta_2(t)=0$, to assure the solutions' decaying when $(x^2+y^2)^{\frac{1}{2}}\to \infty$~\cite{Yu2016Parabola}. Thus, the solution will become as (For simplification, we set $A(t)=\frac{2}{\rho
}\,b(t)^{\frac{1}{4}}\,c(t)^{-\frac{1}{4}}\,\,e^{-\int{\![f(t)-2e_2(t)]dt}}$)
\begin{align}\label{solution}
u&\,=A(t)({\rm
	log} \Phi)_{xx}\, \notag\\
&=A(t) \Big[\frac{b_1 k_1(t){}^2 e^{g_1}+b_3 k_1(t){}^2 e^{-g_1}-b_2 k_2(t){}^2 \cos g_2}{b_1 e^{g_1}+b_3 e^{-g_1}+b_2 \cos
	g_2}-\frac{\left(b_1 k_1(t) e^{g_1}-b_3 k_1(t) e^{-g_1}-b_2 k_2(t) \sin g_2\right){}^2}{\left(b_1 e^{g_1}+b_3 e^{-g_1}+b_2
	\cos g_2\right){}^2}\Big]\,.
\end{align} 

In the following discussion, we set $b(t)=c(t)=d(t)=e_2(t)=\rho=\alpha_1(t)=1$, $f(t)=2$, $\alpha_2(t)=0$ to draw figures. 

\vspace{3mm}
\noindent\textbf{Case 1: Two different parabola characteristic lines}\\
In this case, $C_{k1} C_{k2} \ne 0$ and $C_{k1}/C_{k2}\ne C_{l1}/C_{l2}$, so the two characteristic lines will have the terms of $x$ and $y^2$, with different coefficients.\\

\begin{figure}[H]
	\renewcommand{\captionfont}{\scriptsize}
	\renewcommand{\captionlabelfont}{\scriptsize}
	\centering
	\begin{subfigure}[t]{0.22\textwidth}
		\centering
		\includegraphics[width=1\textwidth]{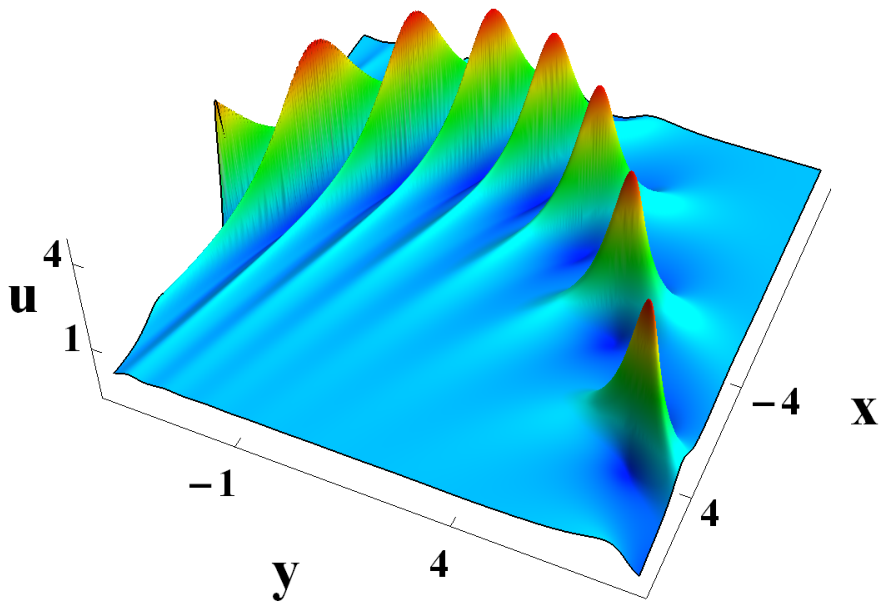}
		\subcaption{}
	\end{subfigure}
	\quad
	\begin{subfigure}[t]{0.22\textwidth}
		\centering
		\includegraphics[width=1\textwidth]{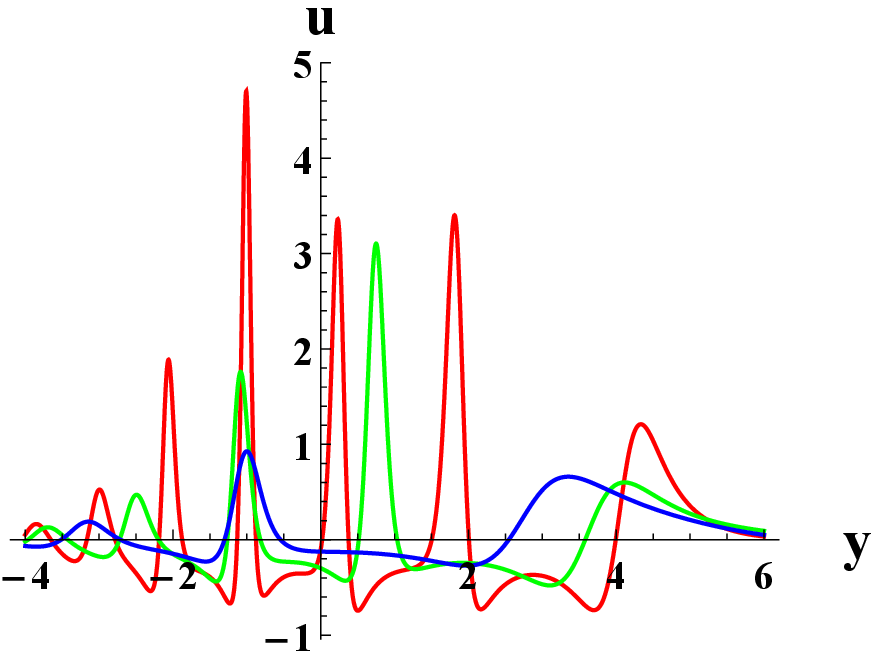}
		\subcaption{}
	\end{subfigure}
	\quad
	\begin{subfigure}[t]{0.22\textwidth}
		\centering
		\includegraphics[width=0.8\textwidth]{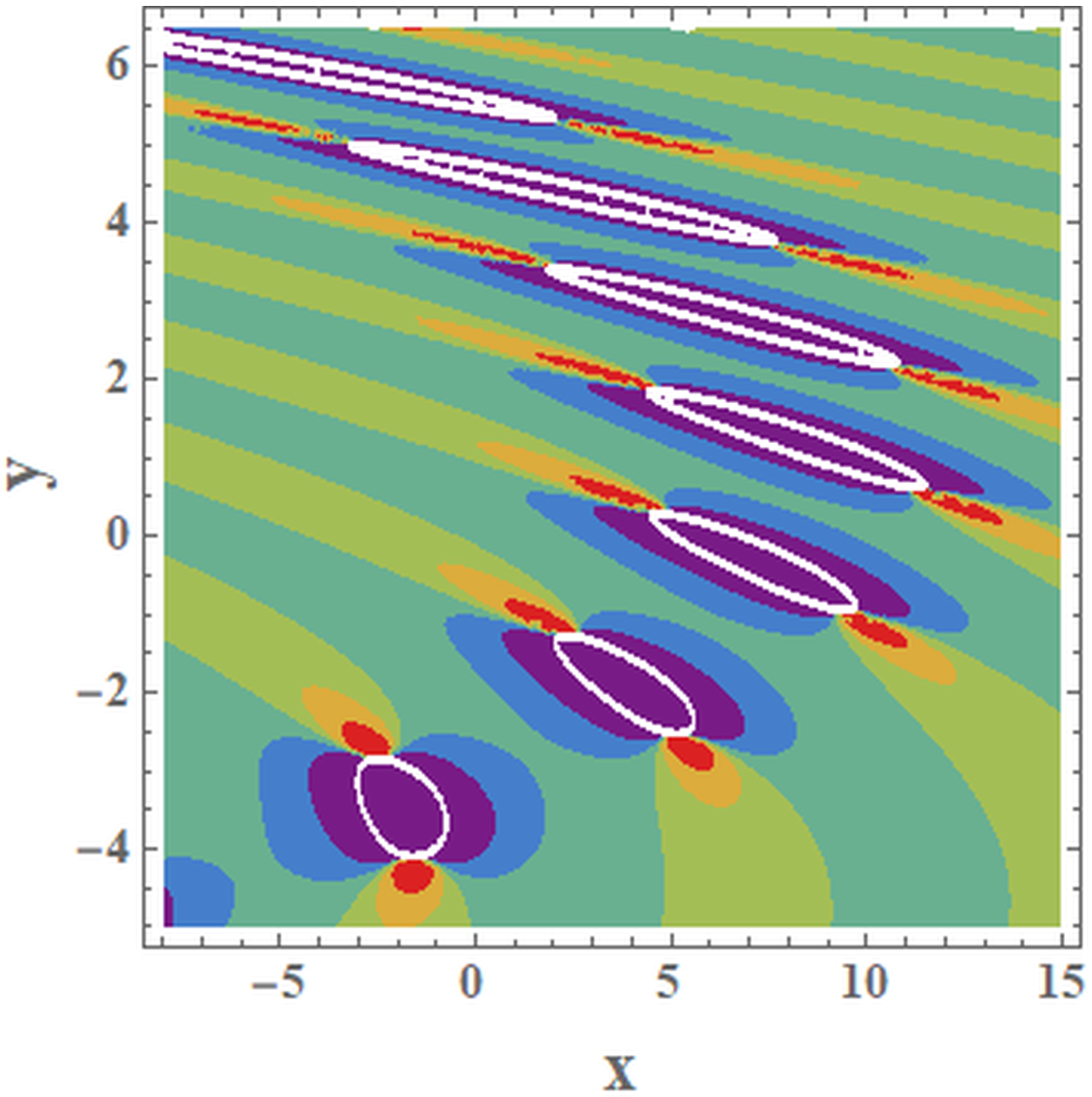}
		\subcaption{}
	\end{subfigure}
	\quad
	\begin{subfigure}[t]{0.22\textwidth}
		\centering
		\includegraphics[width=1\textwidth]{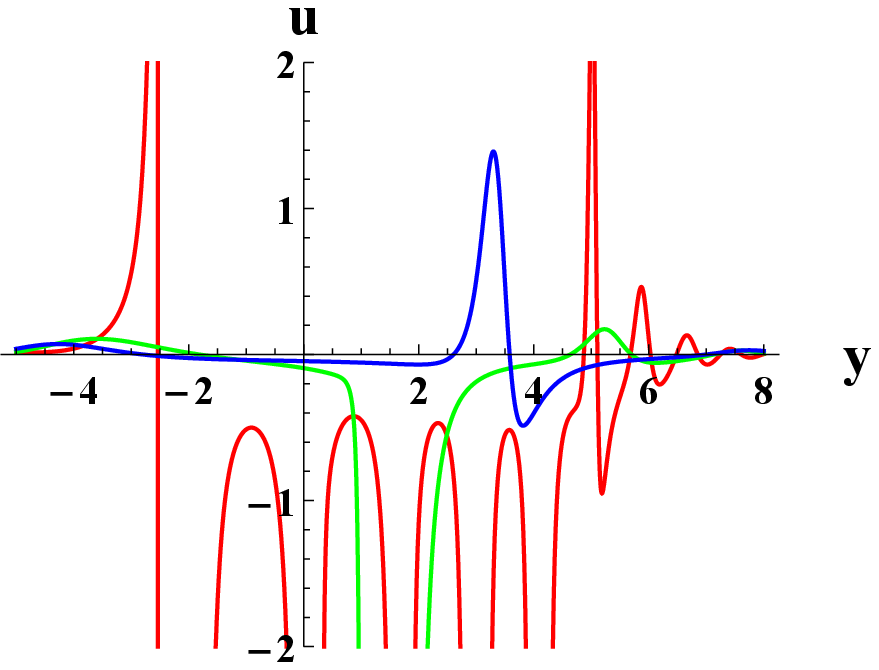}
		\subcaption{}
	\end{subfigure}
	\caption{(a) Solitonic surface for $u$ in Case 1 when $\sigma=-1$, $t=0$.  (b) Profile of the soliton shown in (a) at $x=-12.5$ when $t=0$ (red line), $t=0.2$ (green line) and $t=0.5$ (blue line). (c) Contour plot for $\arctan u$ in Case 1 when $\sigma=1$, $t=0$, where solitons diverge at the white line. (d) Profile of the soliton shown in (c) at $x=5$ when $t=0$ (red line), $t=0.8$ (green line) and $t=1.2$ (blue line). }
	\label{Case1}
\end{figure}

With expression~(\ref{solution}), the solution can be regarded as an interaction between the periodic part and the exponent part. When we just set $b_2=0$, the solution will only have one peak along a parabola. Therefore, for expression~(\ref{solution}), we can consider the trigonometric function offering some impact in some parallel parabolas. For example, when we set $C_{k1}=0.25$, $C_{k2}=0.7$, $C_{l1}=0$, $C_{l2}=4$, $b_2=4$, $b_3=1$, the solitons will be shown as in Fig.~\ref{Case1}: (When dealing with divergence in contour plots, we analyze $\arctan u$ instead to show the features more clearly. Such method is utilized in other cases.)

Modifying the parameters can change the shape of solutions to some extent, including offering some symmetry. For KP equations with $m=C_{l1}/C_{k1}$, $l_{11}(t)$ will be 0 according to Eq.~(\ref{yfangxiang2}). The peak formed by the exponent part will be at a parabola symmetry upon $y=0$.

\begin{figure}[H]
	\renewcommand{\captionfont}{\scriptsize}
	\renewcommand{\captionlabelfont}{\scriptsize}
	\centering
	\begin{subfigure}[t]{0.22\textwidth}
		\centering
		\includegraphics[width=1\textwidth]{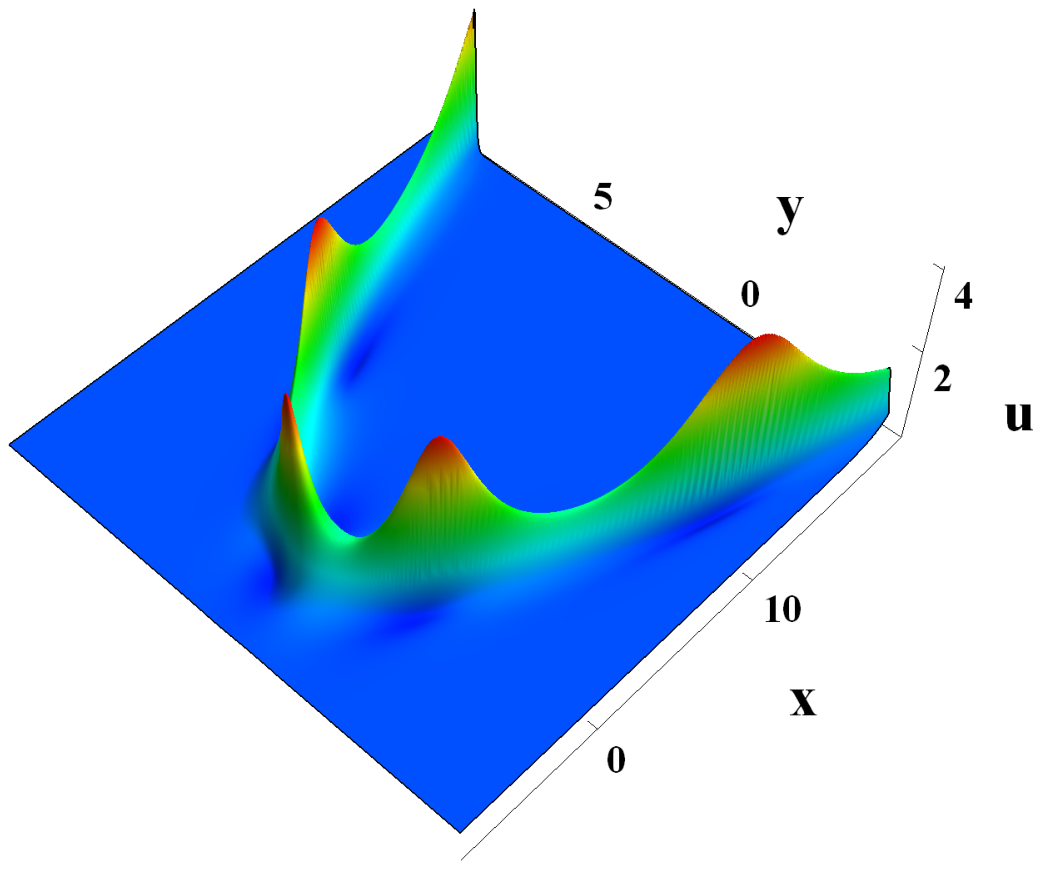}
		\subcaption{}
	\end{subfigure}
	\quad
	\begin{subfigure}[t]{0.22\textwidth}
		\centering
		\includegraphics[width=1\textwidth]{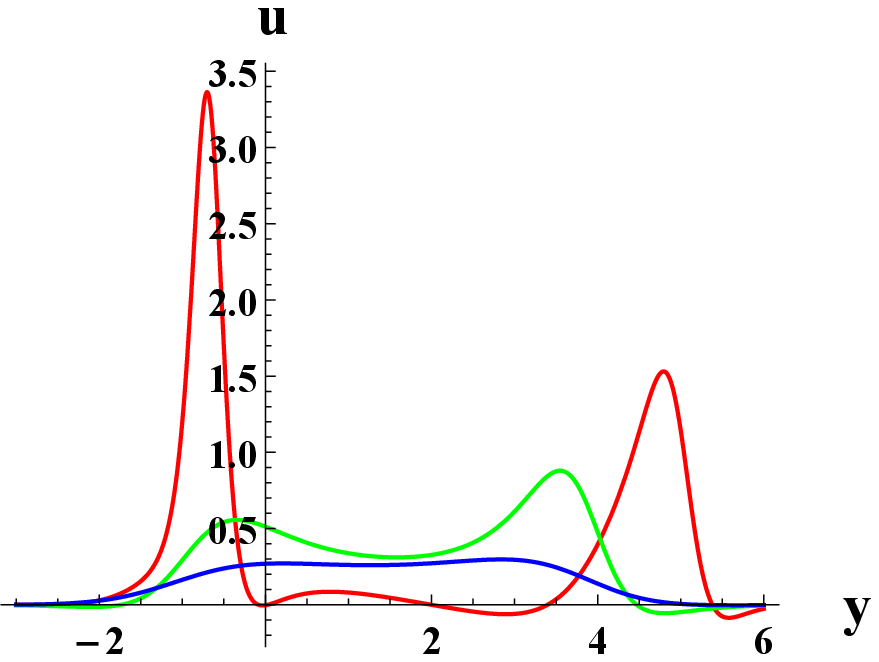}
		\subcaption{}
	\end{subfigure}
	\quad
	\begin{subfigure}[t]{0.22\textwidth}
		\centering
		\includegraphics[width=0.8\textwidth]{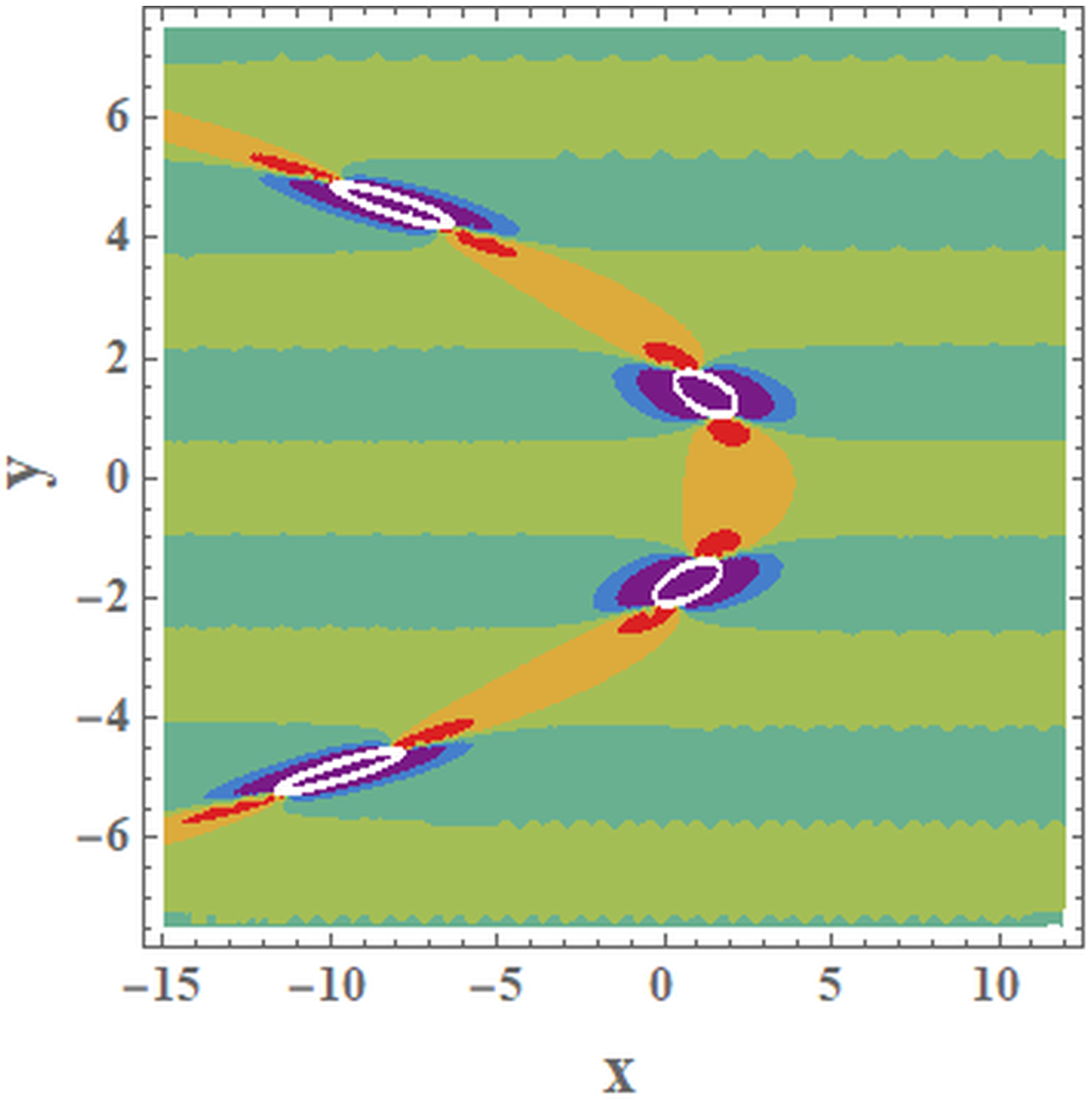}
		\subcaption{}
	\end{subfigure}
	\quad
	\begin{subfigure}[t]{0.22\textwidth}
		\centering
		\includegraphics[width=1\textwidth]{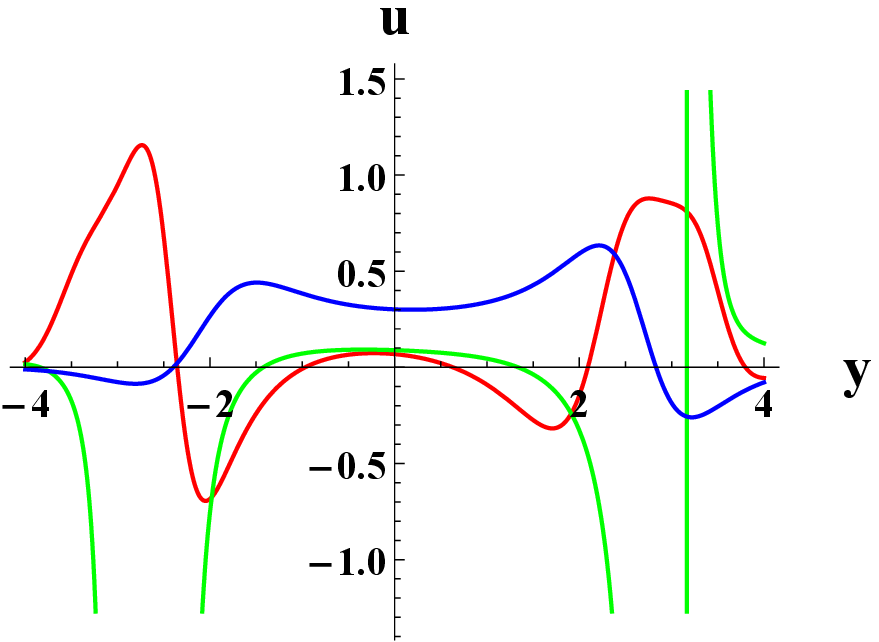}
		\subcaption{}
	\end{subfigure}
	\caption{(a) Solitonic surface for $u$ in Case 2 when $\sigma=-1$, $t=0$.  (b) Profile of the soliton shown in (a) at $x=1$ when $t=0$ (red line), $t=0.5$ (green line) and $t=0.8$ (blue line). (c) Contour plot for $\arctan u$ in Case 2 when $\sigma=1$, $t=0$, where solitons diverge at the white line. (d) Profile of the soliton shown in (c) at $x=-2$ when $t=0$ (red line), $t=0.25$ (green line) and $t=0.5$ (blue line). }
	\label{Case2}
\end{figure}

\vspace{3mm}
\noindent\textbf{Case 2: Straight characteristic line for trigonometric function part}\\
If $C_{k2}=0$, $g_2$ will turn to be a straight line perpendicular to the $x$-axis as $l_{21}(t)y+\omega_2(t)$. Meanwhile, the solution  becomes as
\begin{equation}
u=A(t) \Big[\frac{b_1 k_1(t){}^2 e^{g_1}+b_3 k_1(t){}^2 e^{-g_1}}{b_1 e^{g_1}+b_3 e^{-g_1}+b_2 \cos
	g_2}-\frac{\left(b_1 k_1(t) e^{g_1}-b_3 k_1(t) e^{-g_1}\right){}^2}{\left(b_1 e^{g_1}+b_3 e^{-g_1}+b_2
	\cos g_2\right){}^2}\Big]\,.
\end{equation} 
Here, the trigonometric term will affect the exponent one in equidistant vertical line. When we set $C_{k1}=1$, $C_{k2}=0$, $C_{l1}=1$, $C_{l2}=2$, $b_1=\frac{1}{3}$, $b_2=1$, $b_3=3$, such solution is demonstrated in Fig.~\ref{Case2} 
In this case, we get $b_1=\frac{b_2^2 \sigma  C_{l2}^2}{4 b_3 \left(3 C_{k1}^4+\sigma  C_{l2}^2\right)}$ due to Eq.~(\ref{CON}) ,\space where $b_2,b_3,C_{k1},C_{l1}$ and $C_{l2}$ are free parameters. When setting $b_1=b_3=\frac{\varepsilon b_2 C_{l2}}{2 \sqrt{C_{l2}^2+3 \text{$\sigma $C}_{k1}^4}}\quad(\epsilon=\pm1)$, we have 
\begin{align}
&\hspace{3mm}\notag \Phi(x,y,t)=\frac{\varepsilon b_2 C_{l2}}{ \sqrt{C_{l2}^2+3 \text{$\sigma $C}_{k1}^4}} \cosh \left[k_1(t) x+l_{12}(t) y^2+l_{11}(t) y+w_1(t)\right]\\
&\hspace{23mm}+b_2 \cos \left[l_{21}(t) y+w_2(t)\right]\,.
\end{align}

When $\sigma=-1$ , the existence condition of solution for KP-II enquation is given by $C_{l2}^2-3 C_{k1}^4>0$. However, it is only need that parameters $C_{l2}, C_{k1} $,satisfy $C_{l2}C_{k1}\ne 0$ for KP-I equation ($\sigma=1$)

When $\sigma=-1$ and $C_{l2}^2-3 C_{k1}^4<0$ ,taking  $b_1=-b_3=\frac{\varepsilon b_2 C_{l2}}{2 \sqrt{3 C_{k1}-C_{l2}^2}}$ ,we can obtain
\begin{align}
&\hspace{3mm}\notag \Phi(x,y,t)=\frac{\varepsilon b_2 C_{l2}}{ \sqrt{3 C_{k1}-C_{l2}^2}} \sinh \left[ k_1(t)x+l_{12}(t) y^2+l_{11}(t) y+w_1(t)\right]\\
&\hspace{23mm}+b_2 \cos \left[l_{21}(t) y+w_2(t)\right]\,.
\end{align}

\vspace{3mm}
\noindent\textbf{Case 3: Straight characteristic line for exponent part}\\
Similar to case 2, if $C_{k1}=0$, $g_1$ will be perpendicular to the $x$-axis as $l_{11}(t)y+\omega_1(t)$, and the solution  is
\begin{equation}
u=A(t) \Big[-\frac{b_2 k_2(t){}^2 \cos g_2}{b_1 e^{g_1}+b_3 e^{-g_1}+b_2 \cos
	g_2}-\frac{\left(b_2 k_2(t) \sin g_2\right){}^2}{\left(b_1 e^{g_1}+b_3 e^{-g_1}+b_2
	\cos g_2\right){}^2}\Big]\,.
\end{equation} 
Contrary to the above cases, now there only exist the trigonometric terms in the numerator, which dominate in the solution. Therefore, we will have peak in parallel parabolas, modulated by the exponent term periodically in $y$-direction. Fig.~\ref{Case3} shows one solution of this case with $C_{k1}=0$, $C_{k2}=1$, $C_{l1}=2$, $C_{l2}=1$, $b_2=1$, $b_3=0.5$:
\begin{figure}[H]
	\renewcommand{\captionfont}{\scriptsize}
	\renewcommand{\captionlabelfont}{\scriptsize}
	\centering
	\begin{subfigure}[t]{0.22\textwidth}
		\centering
		\includegraphics[width=1\textwidth]{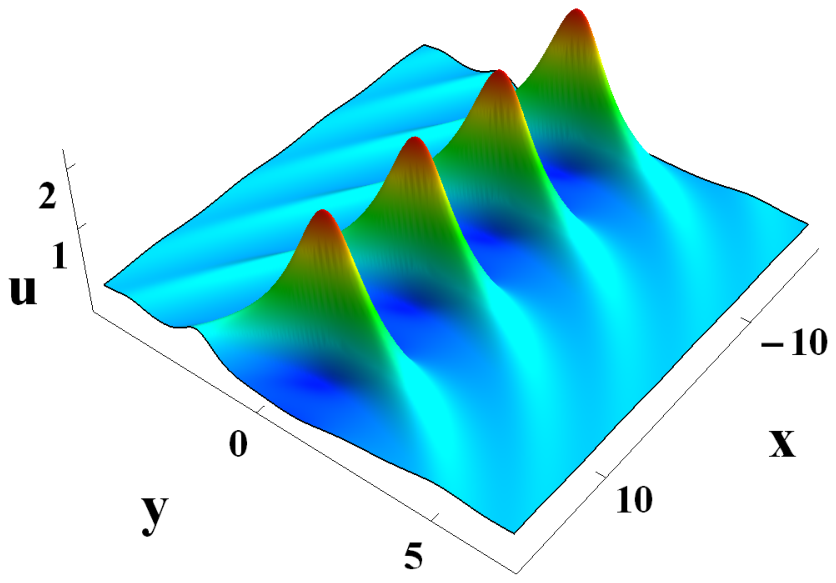}
		\subcaption{}
	\end{subfigure}
	\quad
	\begin{subfigure}[t]{0.22\textwidth}
		\centering
		\includegraphics[width=1\textwidth]{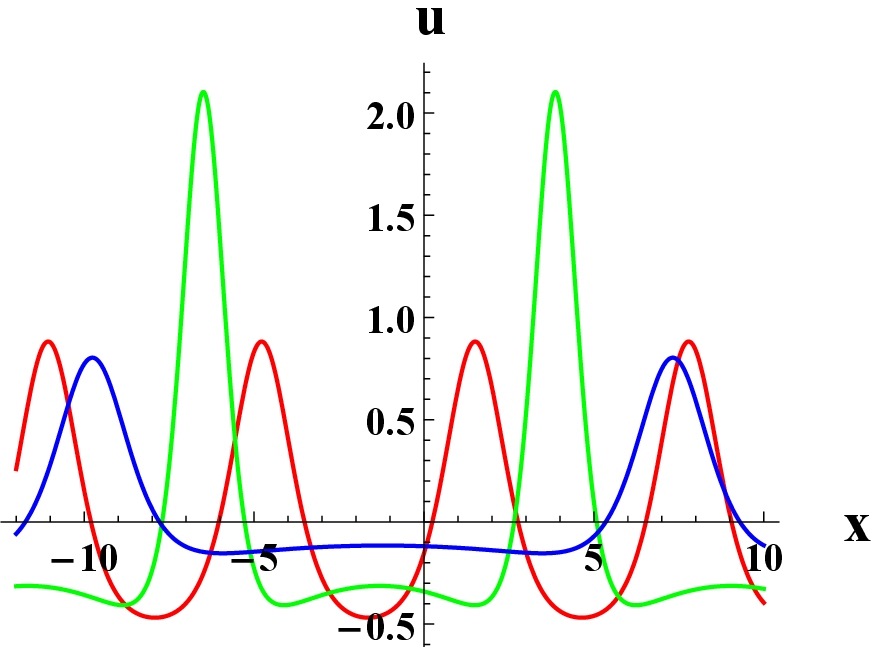}
		\subcaption{}
	\end{subfigure}
	\quad
	\begin{subfigure}[t]{0.22\textwidth}
		\centering
		\includegraphics[width=0.8\textwidth]{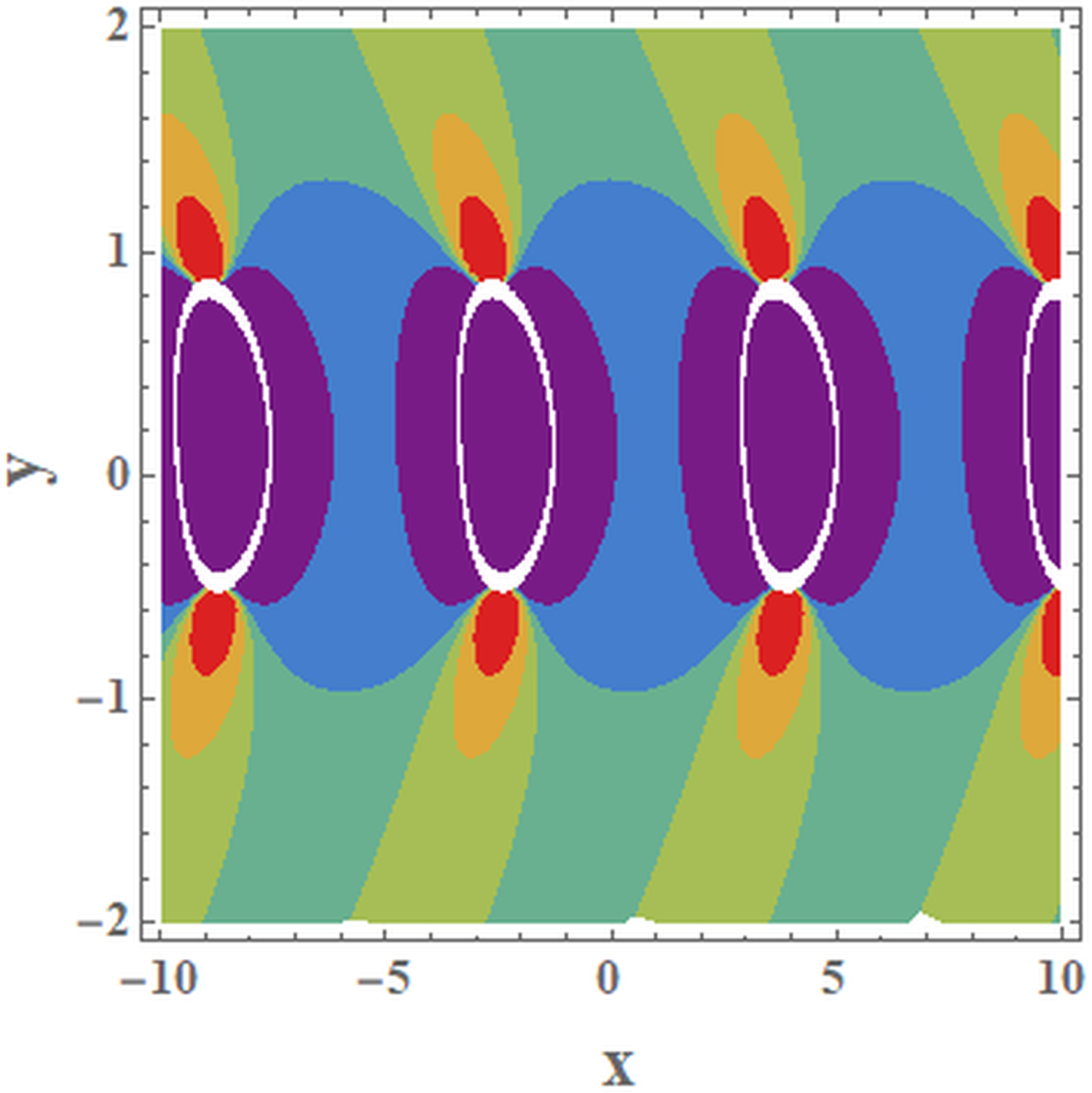}
		\subcaption{}
	\end{subfigure}
	\quad
	\begin{subfigure}[t]{0.22\textwidth}
		\centering
		\includegraphics[width=1\textwidth]{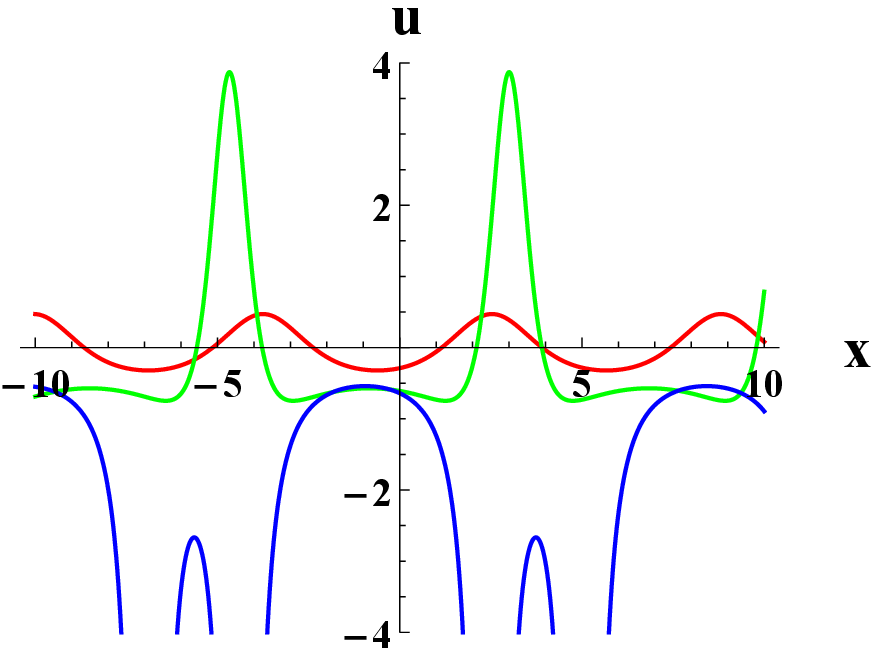}
		\subcaption{}
	\end{subfigure}
	\caption{(a) Solitonic surface for $u$ in Case 3 when $\sigma=-1$, $t=0.5$.  (b) Profile of the soliton shown in (a) at $y=0.25$ when $t=0$ (red line), $t=0.5$ (green line) and $t=1$ (blue line). (c) Contour plot for $\arctan u$ in Case 3 when $\sigma=1$, $t=0$, where solitons diverge at the white line. (d) Profile of the soliton shown in (c) at $y=1.7$ when $t=0$ (red line), $t=0.25$ (green line) and $t=0.4$ (blue line). }
	\label{Case3}
\end{figure}

Here, we get  $b_1=\frac{b_2^2 \sigma  C_{l1}^2-3 b_2^2 C_{k2}^4}{4 b_3 \sigma  C_{l1}^2}$ , where $b_2,b_3,C_{k2},C_{l1}$ and $C_{l2}$ are free parameters. When $\sigma=1$ and $3 C_{k2}^2-C_{l1}^2<0$ ,taking  $b_1=-b_3=\frac{b_2 \varepsilon  \sqrt{3 C_{k2}^2-C_{l1}^2}}{2 C_{l1}}$ ,we can obtain
\begin{align}
&\hspace{3mm}\notag \Phi(x,y,t)=\frac{b_2 \varepsilon  \sqrt{3 C_{k2}^2-C_{l1}^2}}{ C_{l1}} \sinh \left[l_{11}(t) y+w_1(t)\right]\\
&\hspace{23mm}+b_2 \cos \left[k_2(t)x+l_{22}(t) y^2+l_{21}(t) y+w_2(t)\right]\,.
\end{align}
If setting $b_1=b_3=\frac{b_2 \varepsilon  \sqrt{C_{l1}^2-3 \text{$\sigma $C}_{k2}^4}}{2 C_{l1}}$ ,we have 
\begin{align}
&\hspace{3mm}\notag \Phi(x,y,t)=\frac{b_2 \varepsilon  \sqrt{C_{l1}^2-3 \text{$\sigma $C}_{k2}^4}}{C_{l1}} \cosh \left[l_{11}(t) y+w_1(t)\right]\\
&\hspace{23mm}+b_2 \cos \left[k_2(t) x+l_{22}(t) y^2+l_{21}(t) y+w_2(t)\right]\,.
\end{align}

When $\sigma=1$ , the existence condition of solution of solution for KP-I enquation is given by $C_{l1}^2-3 C_{k2}^2>0$. However,it is only need that parameters $C_{l1}, C_{k2} $,satisfy $C_{l1}C_{k2}\ne 0$ for KP-II equation ($\sigma=-1$).

\vspace{3mm}
\noindent\textbf{Case 4: Two parabola characteristic lines with the same shape.}\\
This means $C_{k1}/C_{k2}=C_{l1}/C_{l2}=K$, we get  $b_1=-\frac{b_2^2 k^2}{4 b_3}$ by Eq.~(\ref{CON}) , where $b_2,b_3,K,C_{k1}$ and $C_{l1}$ are free parameters. Since $b_1 b_2<0$, the solution must be divergence along one characteristic line. 

Similarly above, if we set $b_1=-b_3=\frac{\varepsilon b_2 k}{2}$ , then we have 
\begin{align}
&\hspace{3mm}\notag \Phi(x,y,t)={\varepsilon b_2 k} \sinh \left[(k_1(t) x+l_{12}(t) y^2+l_{11}(t) y+w_1(t)\right]\\
&\hspace{23mm}+b_2 \cos \left[
Kk_1(t)x+Kl_{12}(t)y^2 +K l_{11}(t) y+w_2(t)\right]\,.
\end{align}

\vspace{3mm}
\noindent\textbf{Case 5: $b_3=0$}\\
Here the solution changes into 
\begin{equation}
\hspace{14mm}u\,=A(t) \Big[\frac{b_1 k_1(t){}^2 e^{g_1}-b_2 k_2(t){}^2 \cos g_2}{b_1 e^{g_1}+b_2 \cos
	g_2}-\frac{\left(b_1 k_1(t) e^{g_1}-b_2 k_2(t) \sin g_2\right){}^2}{\left(b_1 e^{g_1}+b_2
	\cos g_2\right){}^2}\Big]\,.
\end{equation}
A little different with above, in this case the implicit restraint Eq.~(\ref{CON}) will be:
\begin{equation}\label{neweq}
\hspace{10mm}\ 3 C_{k2}^2 \left(C_{k1}^2+C_{k2}^2\right){}^2=\sigma \left(C_{k2} C_{l1}-C_{k1}
C_{l2}\right){}^2\,,
\end{equation}
which is obvious $\sigma$ must be 1 here because both terms of this equation is positive. Therefore, periodic parabola solutions with $b_3=0$ only appears in KP-I equation. In Figs.~\ref{Case5}(a) and ~\ref{Case5}(b), we set $C_{k1}=0.5$, $C_{k2}=0.5$, $C_{l1}=4 \left(\frac{1}{2}-\frac{\sqrt{3}}{8}\right)$, $C_{l2}=2$, $b_1=3$, $b_2=1$.\\

According to Eq.~(\ref{neweq}), we will find $C_{k1}
C_{l2}=0$ when $C_{k2}= 0$. Furthermore,$C_{k1}=0$ will lead to straight characteristic line solution, and $C_{l2}=0$ will make
$g_2$ as trivial $\omega_2(t)$. As a result, here the characteristic line of trigonometric function mush be parabola.\\

Considering $C_{k1}=0$ with Eq.~(\ref{neweq}) we can get $C_{k2}=\frac{\epsilon  \sqrt{|C_{l1}|}}{\sqrt[4]{3}}$, 
$g_1$ will turn to be a straight line perpendicular to the $x$-axis as $l_{11}(t)y+\omega_1(t)$. Meanwhile, the solution  becomes as
\begin{equation}
u=A(t) \Big[\frac{b_2 k_2(t) \sin
	g_2}{b_1 e^{g_1}+b_2 \cos
	g_2}-\frac{\left (b_2 k_2(t) \sin
	g_2\right){}^2}{\left(b_1 e^{g_1}+b_2
	\cos g_2\right){}^2}\Big]\,.
\end{equation} 
Because $g_1$ turns to be linear as $l_{11}(t)y+\omega_1(t)$, the exponent term will adjust the amplitude of the periodic wave in $y$-direction. When $C_{k1}=0.5$, $C_{l1}=-\frac{\sqrt{3}}{4}$, $C_{l2}=2$, $b_1=3$, $b_2=1$, the solution is like Figs.~\ref{Case5}(c) and ~\ref{Case5}(d):
\begin{figure}[H]
	\renewcommand{\captionfont}{\scriptsize}
	\renewcommand{\captionlabelfont}{\scriptsize}
	\centering
	\begin{subfigure}[t]{0.22\textwidth}
		\centering
		\includegraphics[width=0.8\textwidth]{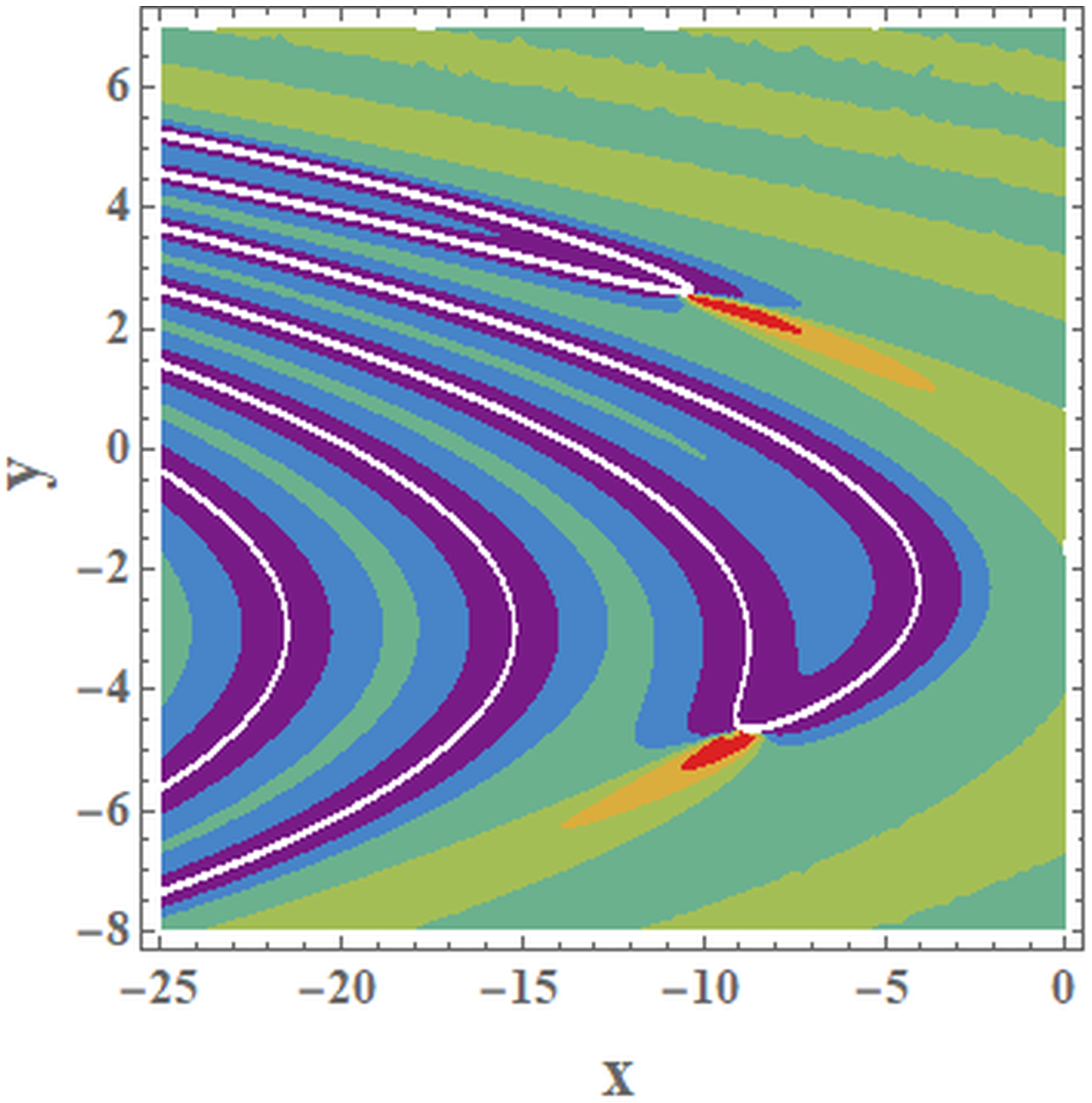}
		\subcaption{}
	\end{subfigure}
	\quad
	\begin{subfigure}[t]{0.22\textwidth}
		\centering
		\includegraphics[width=1\textwidth]{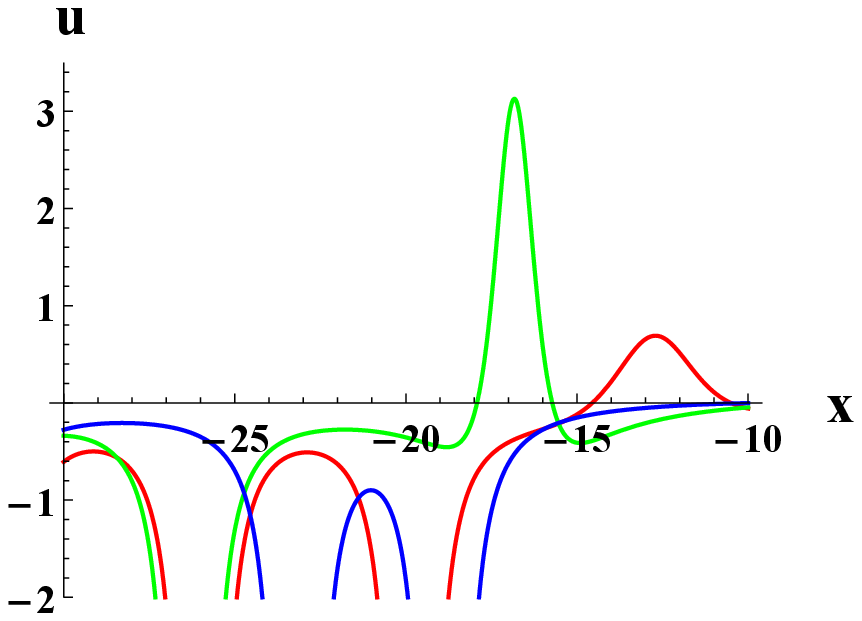}
		\subcaption{}
	\end{subfigure}
	\quad
	\begin{subfigure}[t]{0.22\textwidth}
		\centering
		\includegraphics[width=0.8\textwidth]{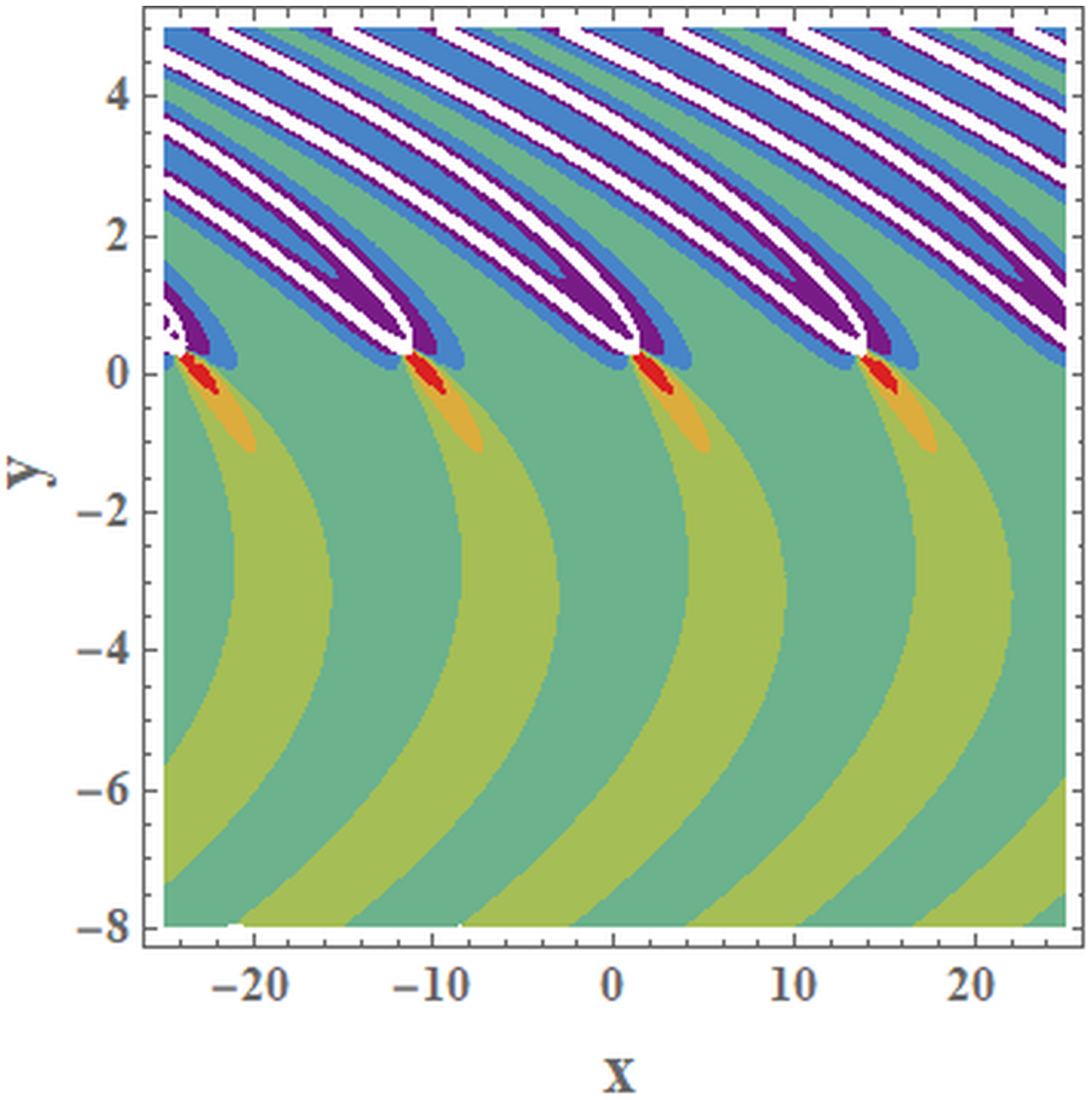}
		\subcaption{}
	\end{subfigure}
	\quad
	\begin{subfigure}[t]{0.22\textwidth}
		\centering
		\includegraphics[width=1\textwidth]{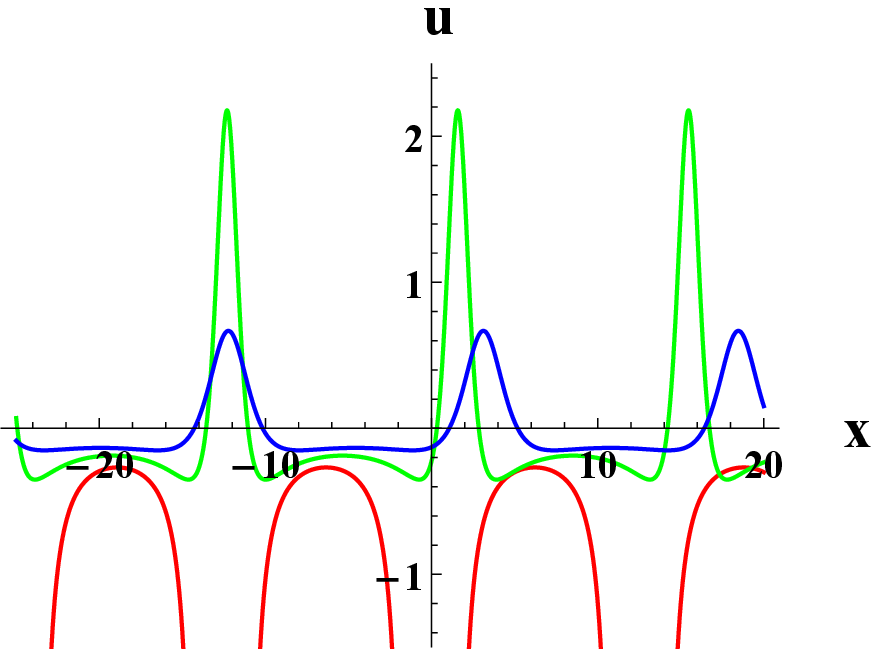}
		\subcaption{}
	\end{subfigure}
	\caption{(a)  Contour plot for $\arctan u$ with both parabola characteristic lines in Case 5 when $t=0$, where solitons diverge at the white line.  (b) Profile of the soliton shown in (a) at $y=-6$ when $t=0$ (red line), $t=0.2$ (green line) and $t=0.4$ (blue line). (c) Contour plot for $\arctan u$ with linear characteristic line for exponent part in Case 5 when  $t=0$, where solitons diverge at the white line. (d) Profile of the soliton shown in (c) at $y=0.7$ when $t=0$ (red line), $t=0.1$ (green line) and $t=0.2$ (blue line). }
	\label{Case5}
\end{figure}

\vspace{3mm}
\noindent {\Large{\bf V. Conclusions}}

\vspace{3mm}
Based on Painlev\'{e} analysis and Hirota bilinear method, periodic parabola solitons for nonautonomous (2+1) dimensional KP equation are obtained. The eleven undetermined parametric functions of periodic parabola solitons are limited to six independent coefficients with one implicit constraint and eight explicit ones. The condition of the solitons' convergence is also found as KP-II equation of different characteristic lines while $b_1 b_3>0$. Besides, five typical cases, classified upon the shape of characteristic lines of the solutions, are discussed and illustrated in this paper, which may appeal to various physics models. Here, all of the results base on the real coefficients. If complex coefficients were considered, we could discuss whether features of the solitons are more complicated and worthy classifying into more cases.

\vspace{3mm}
\noindent {\Large{\bf VI. Acknowledgements}}

\vspace{3mm}
This work has been supported by the National Natural Science Foundation of China under Grant No. 11302014, and by the Fundamental Research Funds for the Central Universities under Grant Nos. 50100002013105026 and 50100002015105032 (Beihang University).

\end{document}